\newcommand{\be}{\begin{equation}}
\newcommand{\ee}{\end{equation}}

\documentstyle[epsf]{article}
\def\boxit#1{\kern4pt\vbox{\hrule\hbox{\vrule\kern8pt\vbox{\kern8pt#1\kern8pt}\kern8pt\vrule}\hrule}}
\def\comma{ \hspace{2mm}, }
\def\period{ \hspace{2mm}. }
\def\and{ {\rm and} }
\def\s{ \hspace{1mm} }
\def\b{ \hspace{2mm} }

\def\de{ {\rm d} }
\def\half{ \frac{1}{2} }
\def\minus{ \mbox - \, }
\def\const{ 8\,\pi\,G }
\def\div{ \partial_v }
\def\dir{ \partial_r }
\def\dirr{ \partial_{rr} }

\begin{document}
\onecolumn

\begin{flushright}
  WATPHYS TH-94/08 \\ gr-qc/yymmddd
\end{flushright}

\vfill

\begin{center}
{\Large \bf Inflationary Behaviour in Axial-symmetric Gravitational Collapse}
\\
\vfill
J.S.F. Chan$^{(1)}$, and R.B. Mann$^{(1,2,3)}$ \\
\vspace{2cm}
(1) Department of Applied Mathematics,
University of Waterloo, Waterloo, Ontario, Canada N2L 3G1\\
(2) Department of Applied Mathematics and Theoretical Physics,
Cambridge University, Cambridge, U.K. CB2 9EW\\
(3) Department of Physics, University of Waterloo, Waterloo,
Ontario, Canada N2L 3G1\\
\vspace{2cm}
PACS numbers: 97.60Lf,04.70.-s,04.20.Dw\\
\end{center}

\vfill

%
%
\begin{abstract}
  We show that the interior of a charged, spinning black hole formed
  from a general axially symmetric gravitational collapse is unstable to
inflation
  of both its mass and angular momentum parameters. Although our
  results are formulated in the context of $(2+1)$-dimensional black
  holes, we argue that they are applicable to $(3+1)$ dimensions.
\end{abstract}

\vfill
\clearpage

\twocolumn

The ultimate fate of a gravitationally collapsing body remains a
subject of considerable interest. The spacetime exterior to a
collapsing body relaxes to that of a Kerr-Newman (KN) black hole
(provided the plausible hypothesis of cosmic censorship holds) with
radiative perturbations decaying as advanced time increases according
to a power law. Infalling matter will either enter a region of
diverging spacetime curvature (where the effects of quantum gravity
are expected to dominate) or avoid this region and emerge via a
`white hole' into another universe, as can occur in the KN case,
where the matter necessarily passes through the Cauchy horizon on
the way. Since it has been shown that this surface is unstable due
to the divergence of the stress-energy of massless test fields
\cite{Penrose}, the question of the final fate of infalling matter
crucially depends upon whether or not this instability seals off
the Cauchy horizon.

New light was shed on this issue when it was demonstrated that in
the presence of outflow from a collapsing body in the Reissner-Nordstr\"{o}m
geometry, the gravitational mass parameter (and thus the curvature
of spacetime) must tend to infinity at the Cauchy horizon, a phenomenon
called mass inflation \cite{Poisson}. Ori subsequently computed the
rate of growth of mass and curvature in a simpler model; from this
he argued that the tidal effects due to spacetime curvature yield
bounded distortions at the Cauchy horizon and so any object attempting
to cross it will not necessarily be destroyed \cite{Ori}. It is possible
that quantum effects substantially modify this picture \cite{quantum},
but the details of this are far from clear.

The restrictive features of these models have inspired theorists to
examine other simpler models, both in $(2+1)$ \cite{Husain} and
$(1+1)$ dimensions \cite{Droz,Balb,JR}. In these models, the
phenomenon is strikingly similar to the Reissner-Nordstr\"{o}m
archetype. However, these studies have all been restricted to the
case where the black holes are non-rotating, which is not general
enough in practice (although suggestive arguments have been given
\cite{BPI,Ori2}). Progress on this front was recently made when
mass inflation was also shown to take place in a $(2+1)$-dimensional
black hole with constant angular momentum \cite{JKR} and a rotating
black string in $(3+1)$ dimensions \cite{BS}. This provides the
strongest evidence to date in support of the hypothesis that mass
inflation will occur in a rotating black hole in General Relativity
\cite{Bonnano}.

In this letter we shall show that the energy parameter of the black
hole (which is dependent upon its mass and angular momentum) diverges
at the Cauchy horizon as a consequence of the interaction between
infalling matter and radiative outflow from a collapsing body. Both
the mass and angular momentum parameters of the black hole can
diverge at the Cauchy horizon -- we argue that in physically
realistic cases the former must diverge faster than the latter.
Although we work in the context of the $(2+1)$ dimensional black
hole \cite{BTZ}, we argue that these features should carry over to
$(3+1)$ dimensions.

Einstein's equations with cosmological constant $\Lambda < 0$ in
$(2+1)$ dimensions
\begin{eqnarray}
  G_{\mu \nu} + \Lambda\,g_{\mu \nu} & = & \const\,T_{\mu \nu} \label{E1}
\end{eqnarray}
have the following exact solution \cite{BTZ}
$$
  \de s^2 =
  \minus N^2(r)\,\de t^2 + N^{\minus 2}\,\de r^2
  + r^2\,\left[\,N^\phi(r)\,\de t + \de \phi\,\right]^2
$$
in the electrovacuum, where $N^\phi(r) \equiv \minus J / (2\,r^2)$
and $N^2(r) \equiv \minus \Lambda\,r^2 - M + \frac{J^2}{4\,r^2}
- 4\,\pi\,G\,q^2\,\ln\left(\frac{r}{r_o}\right)$. The constants of
integration $q$, $M$ and $J$ are the static charge, mass and angular
momentum of the hole respectively \cite{Cangemi,Jolien} and $r_o$ is
an arbitrary scale constant with dimension of length. If in addition
to the electromagnetic stress-energy tensor, we consider a stress-energy
tensor with energy density $\hat{\rho}$ and angular momentum density
$\hat{\omega}$, that is
\begin{eqnarray*}
  \left[\,{\cal T}_{\mu \nu}\,\right] & = &
  \left[\,\begin{array}{ccc}
  \hat{\rho}(v,r)          & 0 & \minus \hat{\omega}(v,r) \\
  0                        & 0 & 0 \\
  \minus \hat{\omega}(v,r) & 0 & 0
  \end{array}\,\right] \comma
\end{eqnarray*}
we obtain
\begin{eqnarray}
  \de s^2 & = &
  \left[\,\Lambda\,r^2 + m(v) + 4\,\pi\,G\,q^2\,\ln(r / r_o)\,\right]\,\de v^2
  \nonumber \\ && \quad
  + \s 2\,\de v\,\de r - j(v)\,\de v\,\de \theta
  + r^2\,\de \theta^2 \comma \label{E2}
\end{eqnarray}
as an exact solution to (\ref{E1}) \cite{JKR}.
Here $m(v)$ and $j(v)$ satisfy the differential equations
$$
  \frac{\de m(v)}{\de v} = 16\,\pi\,G\,\rho(v) \quad \and \quad
  \frac{\de j(v)}{\de v} = 16\,\pi\,G\,\omega(v)
$$
and
$\hat{\rho}(v,r) = \rho(v) / r + j(v)\,\omega(v) / (2\,r^3)$ and
$\hat{\omega}(v,r) = \omega(v) / r$, as dictated by the conservation
laws, with $\rho$, $j$ and $\omega$ arbitrary functions of $v$.
It is furthermore straightforward to show from (\ref{E2}) that the
null geodesic equations simplify to
\begin{eqnarray}
  \frac{\de}{\de \lambda} \left[\,\frac{2}{\dot{v}(\lambda)}\,\right] & = &
  \dir N^2(v,r) \comma \label{E3}
\end{eqnarray}
where $N^2(v,r) \equiv \alpha(v,r) + j^2(v) / (4\,r^2)$ with
$\alpha(v,r) \equiv \minus g_{vv}$, and
\begin{eqnarray}
  2\,\dot{v}\,\dot{r} & = & N^2(v,r)\,\dot{v}^2 \label{E4}
\end{eqnarray}
so that $v$ is a null ingoing coordinate.

Consider a pulse, $S$, of outgoing null radiation between the Cauchy
and outer horizons in the background spacetime (\ref{E2}). We can
model this by matching two patches of solution (\ref{E2}) with
different $m$ and $j$ along $S$ as shown in Figure~1. We denote the
region enclosed by the ring $S$ as II and its complement as region I.
Each region is characterized by its own $m_a(v_a)$ and $j_a(v_a)$,
where $a$ has value of either 1 or 2 for corresponding region. Since
we assume that the two regions have different masses, the Cauchy
horizon cannot coincide with the inner horizon as shown in Figure~3.
If $j_1 = j_2$, the null ring will rotate at the same pace as the
spacetime in regions I and II. However, when $j_1 \neq j_2$, $S$
will carry intrinsic spin.
\vspace*{0.5cm} \\
\epsfxsize=7.5cm
\boxit{\epsffile{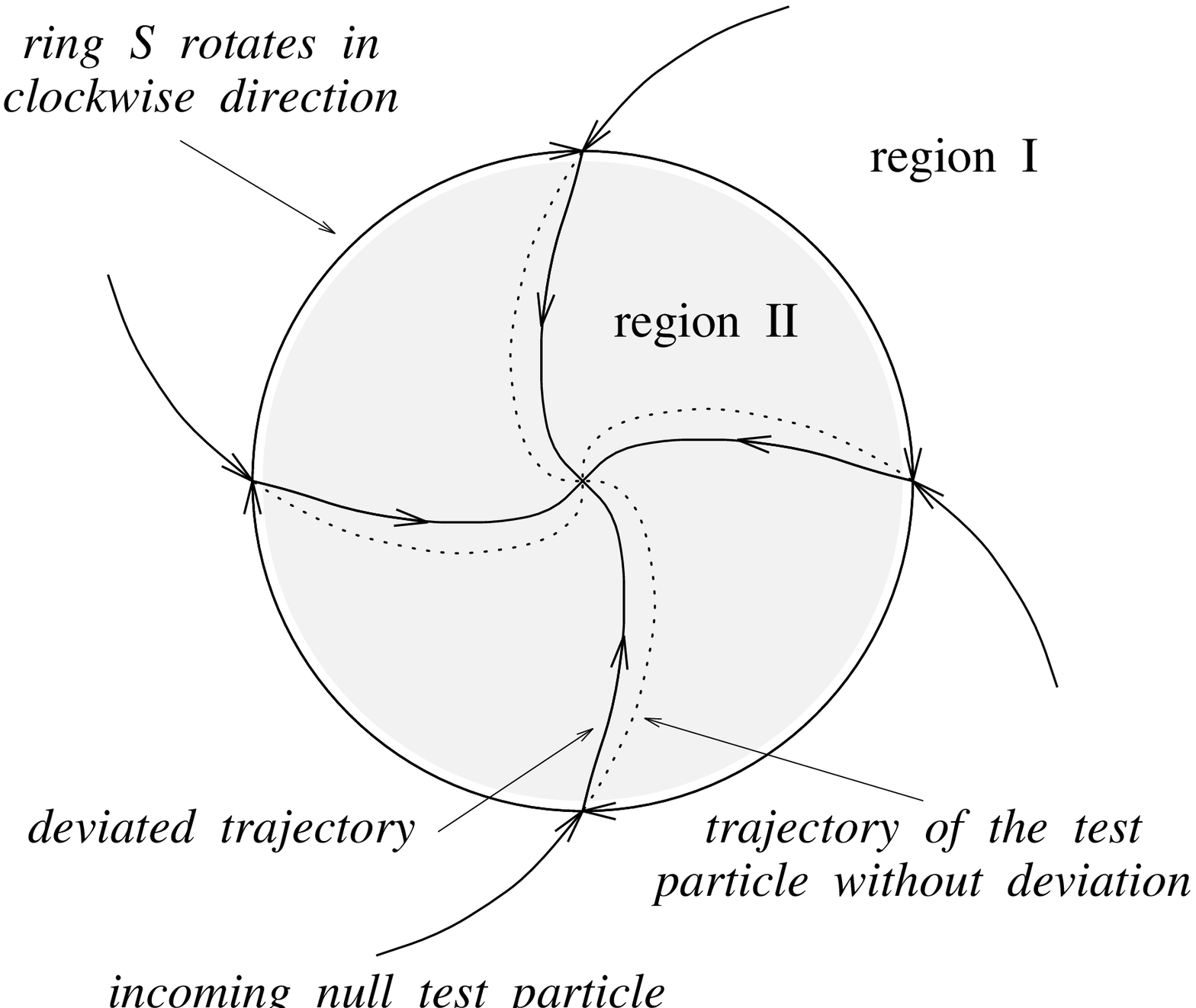}}
\begin{center}
\parbox{6.5cm}
{\small Figure 1: Matching two pieces of non-sta\-tion\-ary BTZ
        solution. Region I has mass $m_1$ and angular momentum $j_1$
	and Region II is characterized by $m_2$ and $j_2$.}
\end{center}
\mbox{\s}

Without loss of generality, we suppose the affine parameter $\lambda$
to be zero at the Cauchy horizon and positive behind that, and in
addition take $M$ and $J$ to be the respective asymptotic values of
$m_1$ and $j_1$. Defining a function $R(\lambda)$ such that $2\,\pi\,R$
is the perimeter of $S$, continuity of inflow along a null curve with
tangent vector
$$
  l_a^{\s \mu} \b = \b
  \left<\,\frac{2}{N_a^{\s 2}} \comma 1 \comma
  \frac{j_a}{r^2\,N_a^{\s 2}}\,\right>
$$
yields (using the null condition (\ref{E4}))
\be
  \frac{\de m_1 - \de (j_1^{\s 2}) / (4\,R^2)}{N_1^{\s 2}} =
  \frac{\de m_2 - \de (j_2^{\s 2}) / (4\,R^2)}{N_2^{\s 2}}
  \period \label{E5}
\ee
When the ring is close to the Cauchy horizon $r_c$, the left side of
equation (\ref{E5}) can be approximated as
\begin{eqnarray*}
  \frac{\de m_1 - \de (j_1^{\s 2}) / (4\,R^2)}{N_1^{\s 2}} & \approx &
  \frac{\minus \de m_1 + \de (j_1^{\s 2}) / (4\,r_c^{\s 2})}
  {\left|\,N_1^{\s 2}(v_1,r_c)\,\right|} \\
  & \approx & \de \ln\left|\,N_1^{\s 2}(v_1,r_c)\,\right|
\end{eqnarray*}
which tends to negative infinity at the Cauchy horizon (where $v_1 \to
\infty$).

The quantity ${E}_a \equiv m_a - j_a^{\s 2} / (4\,r^2)$ appearing
in $N_a^{\s 2}$ is proportional to the total energy of spacetime
at large $r$, neglecting electromagnetic contributions \cite{Jolien}.
It is straightforward to show \cite{prep} that $E$ is positive when
the weak energy condition \cite{Hawking} on null geodesics is
satisfied. One can further show that $E$ is also a geometrically
invariant quantity, since
\begin{eqnarray}
  E & = & \minus g^{\mu \nu}\,\nabla_\mu L\,\nabla_\nu L - \Lambda\,L^2
  \label{E12}
\end{eqnarray}
where $L^2=g(\zeta,\zeta)$ with $\zeta$ a spacelike Killing vector
that obeys the algebra of SO(2) \cite{Lake}. Since the generalized DTR
relation has the same form as in $(3+1)$ dimensions
\cite{BPI}, (except that the dilation rate $K_i$ is half as large), we
find for two colliding null shells
\begin{eqnarray}
  E_D & = & E_A + E_B + \frac{E_A\,E_B}{\Lambda\,r^2} \comma \label{E13}
\end{eqnarray}
where $E_D$ is the energy parameter in region $D$ as shown in Figure~2,
showing that the energy parameters effectively add together for large $r$.
This is the analogue of the $(3+1)$-dimensional result
in ref. \cite{quantum}.
\vspace*{0.5cm} \\
\epsfxsize=7.5cm
\boxit{\epsffile{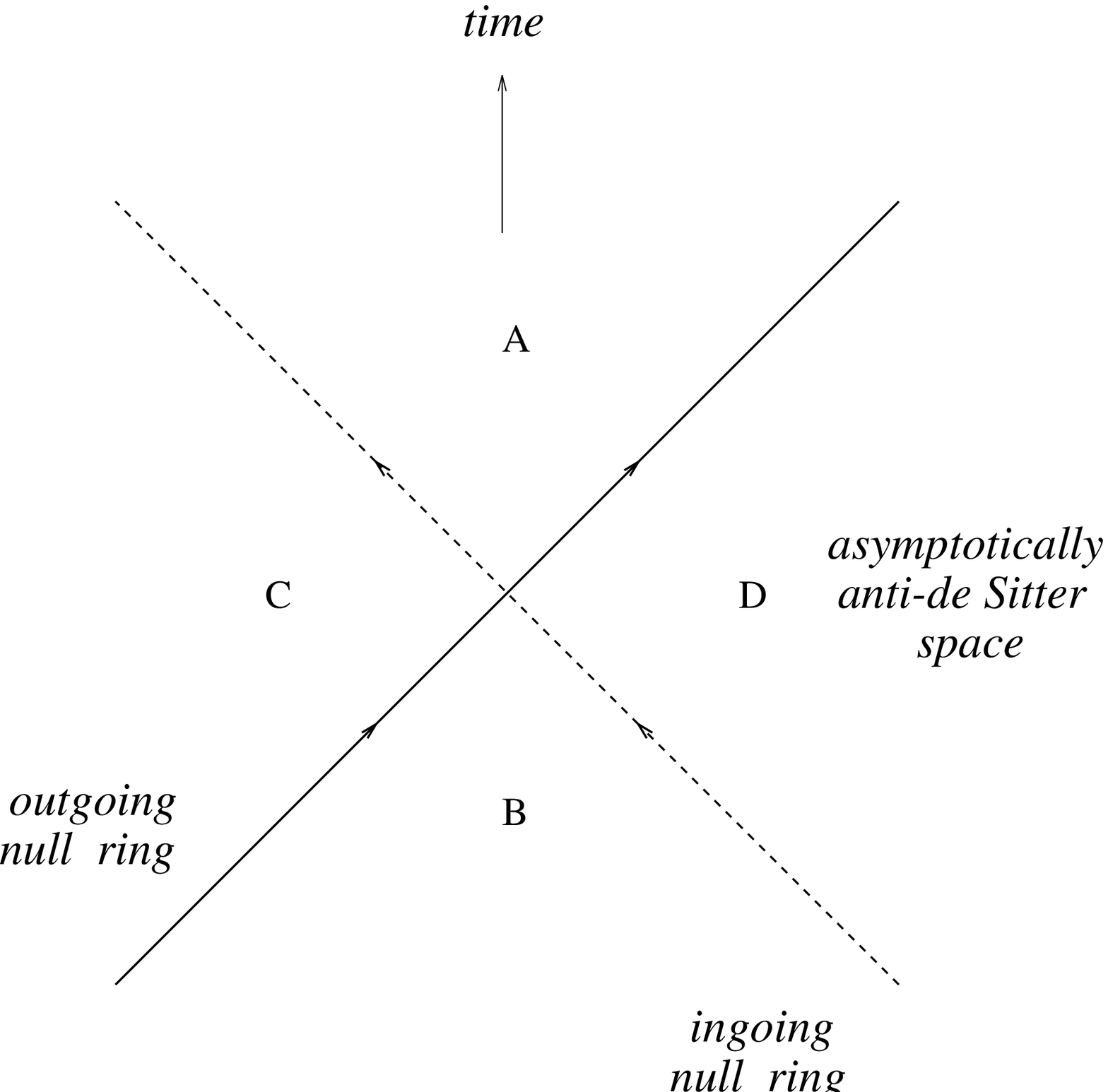}}
\begin{center}
\parbox{6.5cm}
{\small Figure~2: The spacetime diagram for the generalized
	DTR relation. Since the two null rings collide, this event
	divides the spacetime into regions A, B, C and D.}
\end{center}
\mbox{\s}

For constant $q$ the electromagnetic contributions drop out of
(\ref{E5}),
 and so
we shall refer to $E_a$ as the energy parameter in region $a$. The
smaller roots of the transcendental equations $N_a^{\s 2} = 0$ cannot
be the same if the energy content differs between regions; hence if
$S$ has energy $E_{\rm ring} = E_2 - E_1 \neq 0$ \cite{Dray} when
it reaches the Cauchy horizon, $\de m_2 - \de (j_2^{\s 2}) / (4\,r^2)$
will diverge there because $N_2^{\s 2} \neq 0$ at $r_c$. Positivity
of energy in region II (i.e. $m_2 > j_2^{\s 2} / (4\,r^2)$) forces
$m_2$ to have the faster growth rate, forcing the inner horizon to
recede behind the Cauchy horizon as shown in Figure~3.
\vspace*{0.5cm} \\
\epsfxsize=7.5cm
\boxit{\epsffile{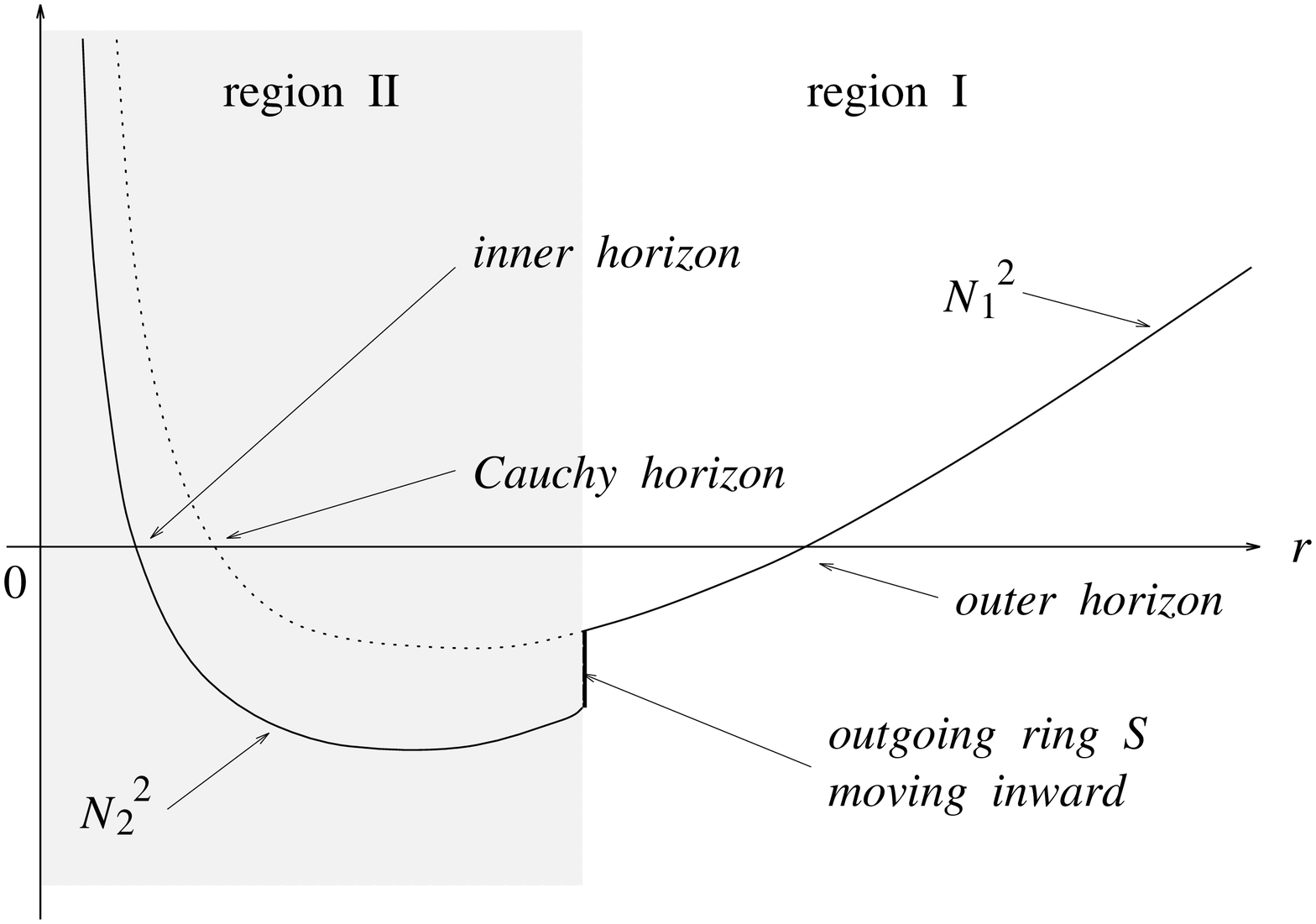}}
\begin{center}
\parbox{6.5cm}
{\small Figure 3: Horizon locations of the matched spacetime.}
\end{center}
\mbox{\s}

We can approximate the rate of energy inflation as follows. From
(\ref{E3}) and (\ref{E4}) we have
\be
  m_a(v_a(\lambda)) - \frac{j_a^{\s 2}(v_a(\lambda))}{4\,R^2(\lambda)} =
  {\cal H}(R(\lambda))
  - \dot{R}(\lambda)\,\frac{z_a(\lambda)}{R(\lambda)} \comma \label{E6}
\ee
\be
  v_a(\lambda) \b = \b
  2\,\int^\lambda \frac{R(\zeta)}{z_a(\zeta)}\,\de \zeta \comma \label{E7}
\ee
\be
  \frac{z_a(\lambda)}{R(\lambda)} = Z_a
  + \int^\lambda_0 \dir \left. N_a^{\s 2}(v_a(\zeta),r) \right|_{r=R(\zeta)}\,
  \de \zeta \comma \label{E8}
\ee
where ${\cal H}(R) \equiv
\minus \Lambda\,R^2 - 4\,\pi\,G\,q^2\,\ln\left(R / r_o\right)$
has the  limit $M - J_1^{\s 2} / (4\,r_c^{\s 2})$ when $R$ tends to
$r_c$ and $Z_a$ are constants. Since $N_1^{\s 2}$ and $N_2^{\s 2}$
have finite slope at the Cauchy horizon, $z_a(\lambda)$ can be
approximated as
\begin{eqnarray*}
  \frac{z_1(\lambda)}{R(\lambda)} \b \approx \b Z_1 - 2\,k_o\,\lambda & \and &
  \frac{z_2(\lambda)}{R(\lambda)} \b \approx \b Z_2 - K\,\lambda
\end{eqnarray*}
for small $\lambda$, where $k_o$ and $K$ are constants. The constant
$k_o$ must be positive since the slope of $N_1^{\s 2}$ at the Cauchy
horizon is always negative. (See Figure~3.) As $\lambda \to 0$ we
expect $\lim_{\lambda \rightarrow 0^{\minus}} \dot{v}_1(\lambda) =
2 / Z_1 = \infty$ but that $\dot{v}_2$ is finite, implying that $Z_1 = 0$
and $Z_2 \neq 0$. Thus
\begin{eqnarray}
  v_1(\lambda) \b \approx \b \minus \frac{1}{k_o}\,\ln|\lambda| & \and &
  v_2(\lambda) \b \approx \b \frac{2}{Z_2}\,\lambda \period
  \hspace{8mm} \label{E9}
\end{eqnarray}
Since $N_1^{\s 2} = \minus |\,{\cal H}(r)-m_1+j_1^{\s 2}/(4\,r^2)\,|
\approx \minus |\,M-J^2/(4\,r_c^{\s 2})-m_1+j_1^{\s 2}/(4\,r_c^{\s 2})\,|$
just outside the Cauchy horizon, we have
$\dot{R}(\lambda) \approx \delta E(\lambda) / (2\,k_o\,\lambda)$
from equation (\ref{E4}), where
$$
  \delta E(\lambda) =
  M - m_1(\lambda) - \frac{J^2}{4\,r_c^{\s 2}}
  + \frac{j_1^{\s 2}(\lambda)}{4\,r_c^{\s 2}}
$$
is the tail of the total energy at late time. Thus equation
(\ref{E6}) implies that the energy of the ring is
\begin{eqnarray*}
  E_{\rm ring}(\lambda) & \approx &
  \minus \frac{Z_2}{2\,k_o\,\lambda}\,\delta E(\lambda) \comma
\end{eqnarray*}
where $Z_2$ must be positive so that $E_{\rm ring} > 0$. If we
assume that $\delta E(\lambda)$ decays to zero via a power law
$h\,v_1^{\s \minus p}$ \cite{JKR,Price}, we obtain
\begin{eqnarray}
  E_2(v_2) & \approx &
  M - \frac{J^2}{4\,r_c^{\s 2}} \nonumber \\ &&
  - \s \frac{h}{v_2}\,k_o^{\s p-1}\,
  \left|\,\ln\left|\frac{Z_2\,v_2}{2}\right|\,\right|^{\minus p}
  \period \label{E10}
\end{eqnarray}
As a result, $E_2(v_2)$ goes to infinity while $S$ approaches the Cauchy
horizon because $v_2$ tends to zero from below at that instant.

We have shown that the phenomenon of mass inflation is replaced by
the more general phenomenon of inflation of the energy parameter in
the case of collapse of a null fluid with non-zero angular momentum
density. Although this conclusion arises from an analysis of
$2+1$-dimensional black holes, we expect that the qualitative
features of this phenomenon (namely that both mass and angular
momentum parameters inflate) carry over to $3+1$ dimensions for
several reasons. First of all, the results of this paper carry over
straightforwardly to the case of a charged, spinning black string
in $(3+1)$ dimensions \cite{BS,prep}. Second, the result (\ref{E10})
for constant angular momenta is qualitatively similar to the situation
hypothesized for axisymmetric collapse to a Kerr black hole \cite{BPI}
-- as the mass parameter inflates, the angular momentum per unit mass
$J/m$ becomes negligible. We expect that when fluids with non-vanishing
angular momentum density are included, the angular momentum parameter
will inflate in a manner consistent with the results of ref. \cite{Ori2}.

It is also possible to consider the results of a fluid with vanishing
$\rho(v)$, yielding constant $m$'s in each of regions I and II. We find
that the growth rate of $j_2^2$ is qualitatively similar to that of
$E_2$ in (\ref{E10}) but is opposite in sign, diverging to negative
infinity. This unphysical situation is associated with the breakdown
of the positivity of energy due to the vanishing of $\rho(v)$ \cite{prep}.

We close by considering tidal distortions at the horizon. In the
triad frame, the relevant components of the Riemann tensor look
like the following:
\begin{eqnarray*}
  R^1{}_{001} & = &
  \frac{\div \alpha - \dir \alpha}{2\,r} - \frac{\div (j^2)}{8\,r^3}
  - \frac{j^2}{8\,r}\,\dir \left(\,\frac{\dir \alpha}{r}\,\right) \comma \\
  R^1{}_{002} & = & R^2{}_{001} \b = \b
  \frac{j}{4}\,\dir \left(\,\frac{\dir \alpha}{r}\,\right)
  + \frac{\div j}{2\,r^2} \comma \\
  R^2{}_{002} & = & \minus \half\,\dirr \alpha \period
\end{eqnarray*}
It is clear that the most divergent component is $R^1{}_{001}$.
The tidal distortion is finite since one
can approximate the distortion by integrating the above components
twice with respect to $v$ \cite{Ori,Balbinot} and obtain a finite
result.
Furthermore the Kretschmann scalar of the BTZ solution is given by
$$
  R_{abcd}\,R^{abcd} =
  \frac{2}{r^2}\,\left[\,\dir \alpha(v,r)\,\right]^2
  + \left[\,\dirr \alpha(v,r)\,\right]^2
$$
which is obviously bounded at Cauchy horizon. As a result, we see no
reason to terminate the classical extension of the spacetime beyond
the Cauchy horizon, as Ori has suggested \cite{Ori,Ori2}.

\bigskip

This work was  supported by the Natural Sciences and Engineering
Research Council of Canada.

\end{document}